\input harvmac
\input epsf

\def\p{\partial}


\Title{}{\vbox{{\centerline{Holographic Principle and Quantum
Cosmology}}}}

\centerline{Qing-Guo Huang}

\medskip
\centerline{\it Interdisciplinary Center of Theoretical Studies}
\centerline{\it Chinese Academia Sinica, Beijing 100080, China}
\medskip
\centerline{\it and}
\medskip
\centerline{\it Institute of Theoretical Physics, Chinese Academia
Sincia} \centerline{\it P. O. Box 2735, Beijing 100080, China}
\bigskip

\centerline{\tt huangqg@itp.ac.cn}

\bigskip

\bigskip

Using the holographic entropy proposal for a closed universe by
Verlinde, a bound on equations of state for different stages of
the universe is obtained. Further exploring this bound, we find
that an inflationary universe naturally emerges in the early
universe and today's dark energy is also needed in the quantum
cosmological scenario.


\Date{December 2005}

\nref\wmap{C. L. Bennett et al., Astrophys.J.Suppl. 148 (2003) 1,
astro-ph/0302207; D. N. Spergel et al., Astrophys. J. Suppl. 148
(2003) 175, astro-ph/0302209. }

\nref\inf{A. A. Starobinsky, JETP Lett. 30 (1979) 682, Phys. Lett. B
91 (1980) 99; A. H. Guth, Phys. Rev. D 23 (1981) 347; A. Linde,
Phys. Lett. B 108 (1982) 389; A. Albrecht and P. J. Steinhardt,
Phys. Rev. Lett. 48 (1982) 1220. }

\nref\hop{C. R. Stephens, G. 't Hooft and B. F. Whiting,
Class.Quant.Grav. 11 (1994) 621, gr-qc/9310006; L. Susskind,
J.Math.Phys. 36 (1995) 6377, hep-th/9409089. }

\nref\adscft{J. Maldacena, Adv.Theor.Math.Phys. 2 (1998) 231,
hep-th/9711200; E. Witten, Adv.Theor.Math.Phys. 2 (1998) 253,
hep-th/9802150. }

\nref\dscft{A. Strominger, JHEP 0110 (2001) 034, hep-th/0106113; A.
Strominger, JHEP 0111 (2001) 049; Qing-Guo Huang, K. Ke and M. Li,
hep-th/0505020.}

\nref\fs{W. Fischler and L. Susskind, hep-th/9806039. }

\nref\kl{N. Kaloper and A. Linde, Phys. Rev. D 60 (1999) 103509,
hep-th/9904120. }

\nref\rb{R. Bousso, JHEP 9907(1999)004, hep-th/9905177; R. Bousso,
JHEP 9906(1999)028, hep-th/9906022; R. Bousso, Rev. Mod. Phys. 74
(2002) 825, hep-th/0203101. }

\nref\br{D. Bak and S-J. Rey, Class. Quant. Grav. 17 (2000) L83,
hep-th/9902173. }

\nref\elv{R. Easther and D. Lowe, Phys. Rev. Lett. 82 (1999) 4967,
hep-th/9907032; G. Veneziano, Phys. Lett. B 454 (1999) 22,
hep-th/9902126. }

\nref\aas{A. A. Starobinsky, Grav.Cosmol. 6 (2000) 157,
astro-ph/9912054. }

\nref\hh{J. B. Hartle and S. W. Hawking, Phys. Rev. D 28 (1983)
2960. }

\nref\qh{Qing-Guo Huang, hep-th/0510219. }

\nref\ev{E. Verlinde, hep-th/0008140. }

\nref\fl{S. R. Coleman and F. De Luccia, Phys. Rev. D 21 (1980)
3305; A. Linde, JCAP 0410 (2004) 004, hep-th/0408164; B. Mclnnes,
hep-th/0511227. }

\nref\fst{H. Firouzjahi, S. Sarangi and S. H. Henry Tye, JHEP 0409
(2004) 060, hep-th/0406107. }

\nref\st{S. Sarangi and S. H. Henry Tye, hep-th/0505104. }

\nref\bm{B. Mclnnes, hep-th/0509035. }

\nref\ba{R. Brustein and S. P. de Alwis, hep-th/0511093. }

\nref\hm{R. Holman and L. Mersini-Houghton, hep-th/0511102, 0511112,
0512070. }

\nref\ex{A. G. Riess et al., Astron. J. 116 (1998) 1009,
astro-ph/9805201; S. Perlmutter et al., Astrophys. J. 517 (1999)
565, astro-ph/9812133; A. G. Riess et al., Astrophys. J. 607 (2004)
665, astro-ph/0403292. }

Hot big bang model says that our universe was born at some moment
$t=0$ about 13.7 billion years ago \wmap, in a state with
infinitely large energy density and temperature. With the rapid
expansion of the universe the average energy of particles and the
temperature of the universe decreased rapidly and the universe
became cold. This theory have been popularly accepted after the
discovery of the cosmic microwave background radiation. However,
like its counterpart in particle physics, Hot big bang model is
not without its shortcomings, including many intrinsic
difficulties, such as flatness problem, horizon problem, monopoles
problem and so on. They are not inconsistencies within this model
itself; rather, they involve questions that this model in the
splendor of its success allows one to ask, but for which the model
has yet to provide answers.

Fortunately, all these problems can be solved simultaneously in the
inflationary universe scenario \inf. The idea of inflation is that
the expansion of the universe during some period of the early
universe, known as the inflationary stage, is accelerating. This
rapid expansion made the density of monopoles vanishingly small and
the size of the observable universe smaller than the Hubble size
during inflation as well. Usually inflation was driven by the
effective potential of the inflaton field. Nevertheless the quantum
fluctuations of the inflaton field provided the seeds for the
formation of the large scale structure of our universe and the
temperature fluctuations in CMB, which have been confirmed by the
cosmological observations, for instance \wmap. But there is not a
fundamental theory which predicts that inflation must happen in the
early universe.

What is the initial condition for our universe is one of the deepest
questions in modern physics. We believe a well-understood quantum
theory of gravity is needed before we can answer this question. In
the last ten years, many unexpected lessons about the nature of
spacetime have been learned by string theory and black hole theory.
We believe that the concept of holography must be one of the key
concepts for quantum gravity \refs{\hop- \dscft}. Several
conjectures of the holographic principle for cosmology have also
been suggested and many consequences are obtained, for example
\refs{\fs-\elv}. In \refs{\fs, \kl}, the authors found a holographic
bound on the equation of state. But the holographic principle in
\refs{\fs, \kl} fails to describe a closed univere. The covariant
entropy bound proposed by Bousso in \rb\ is still valid for a closed
universe, but the second law of thermodynamics cannot be
responsible. On the other hand, there are a number of conjectures
about the origin of our universe (see \aas\ for a brief review). As
the most attractive idea among them, Hartle-Hawking no-boundary wave
function $\Psi_{HH}$ of the universe \hh\ says that our universe was
born of a tunnelling from nothing. Nothing means a state without any
classical spacetime, or, a state with zero entropy \qh. We expect
that holographic principle and quantum cosmology can provide some
insights on the initial conditions of our universe.

Relying on the holographic description for a closed universe
proposed by Verlinde \ev\ and the second law of thermodynamics, we
get a bound on the equation of state and we find that an
inflationary universe naturally emerges and today's dark energy is
also needed in quantum cosmological scenario.

Let us start with Friedman-Robertson-Walker (FRW) metric for a
$(n+1)$-dimensional closed universe \eqn\metr{ds^2=-dt^2+a^2(t)
d\Omega_n^2, } where $a(t)$ represents the radius of the universe
and $d\Omega_n^2$ is a short hand notation for the metric on the
unit $n$-sphere $S^n$. The spatial volume of this
$(n+1)$-dimensional closed FRW universe is given by
\eqn\vcu{V=\hbox{Vol}(S^n)a^n. } The volume of a closed universe is
finite. In $n+1$ dimensional spacetime the FRW equations are given
by \eqn\fdm{H^2={16 \pi G \over n(n-1)} \rho - {1 \over a^2},}
\eqn\sfdm{\dot H=-{8\pi G \over n-1}(\rho+p)+{1 \over a^2},} where
$H={\dot a /a}$ is the Hubble parameter and the dot denotes as
differentiation with respect to the time $t$. In $(n+1)$ dimensions
the equation of state for radiation is $w=p/\rho=1/n$, for dust-like
matter $w=0$ and for a cosmological constant $w=-1$. Using eq. \fdm\
and \sfdm, we find \eqn\acd{{\ddot a \over a}=H^2+{\dot H}={8\pi G
\over n-1}\left({2 \over n}-1-w \right) \rho. } The expansion of the
universe is accelerating, if \eqn\cac{w<w_c=-1+{2 \over n}.} This
result is also valid for a flat or an open universe.

On the other hand, the holographic principle says that the maximum
entropy in a region of space is \hop\ \eqn\hps{S_{max}={A \over 4G},
} where $A$ is the area of the boundary of the region and $G$ is the
Newton coupling constant. But since the space for a closed universe
has no boundary, the holographic principle in its naive form \hps\
does not work in a closed universe.

Fortunately, Verlinde found a deep relationship between the entropy
formulas for the CFT and the FRW equations for a closed universe. In
\ev, Verlinde focus on the case with a radiation dominated closed
universe. But he also pointed out that the matching of the FRW
equations and the Cardy formula is independent on the equation of
state of the matter. Therefore, from the dual CFT point of view, we
propose that the entropy for a closed universe whose evolution is
dominated by the matter with arbitrary equation of state always
takes the form \eqn\entp{S=(n-1){HV \over 4G}. } With the evolution
of the universe, its entropy also varies. Based on the spirit of the
second law of thermodynamics, it is interesting for us to
investigate the variation of the entropy with time, \eqn\sdt{{d S
\over dt}={S \over H} (n H^2 + {\dot H})={8\pi G \over n-1} {S \over
H} (\rho-p+2{n-1 \over n}\rho_k), } where the energy density of the
curvature $\rho_k$ is defined as \eqn\dc{\rho_k=-{n(n-1) \over 16\pi
G} {1 \over a^2}. } The energy density of the curvature is negative
for a closed universe, positive for an open universe and equals zero
for a flat universe. The equation of state for the curvature energy
density is $w_k=-1+2/n$ which is just the borderline between
decelerated and accelerated expansion. Physically the entropy can
not be negative, because it is proportional to $\ln {\cal N}$, where
${\cal N}$ is the number of states for a system and should be a
positive integer. An interesting result comes from the second law of
the thermodynamics which requires that the entropy of the universe
can not decrease, namely $dS/dt\geq 0$. Thus \eqn\crp{\rho-p+2{n-1
\over n}\rho_k\geq 0, } for an expansive universe, where the Hubble
parameter $H$ is positive. We re-write the Friedman equation \fdm\
as \eqn\fdo{\rho_c=\rho+\rho_k, } or equivalently,
$1=\Omega+\Omega_k$, here \eqn\rc{\rho_c={n(n-1)H^2 \over 16 \pi G}}
is the critical energy density, $\Omega=\rho/\rho_c$ and
$\Omega_k=\rho_k /\rho_c$. Combining eq. \crp\ and \fdo, an upper
bound on the equation of state for the energy density $\rho$ is
obtained \eqn\cw{w={p \over \rho} \leq 1+2 \left( 1-{1 \over n}
\right) {1-\Omega \over \Omega}. } For a flat universe,
$\Omega_k=0$, or $\Omega=1$, eq. \cw\ becomes $w \leq 1$, which is
just the dominant energy condition and can be interpreted as saying
that the speed of energy flow of matter is always less than the
speed of light. This offers an evidence to support our proposal. A
similar result for a flat universe is also obtained in
\refs{\fs-\rb}. However the holographic principle for the universe
proposed in \refs{\fs,\kl} is violated for a closed universe. Even
though the covariant entropy bound proposed by Bousso in \rb\ is
still valid for a closed universe, it is time reversal invariant and
the second law of thermodynamics cannot be responsible. Here we
stress that the starting point of ours is quite different from those
by the authors of \refs{\fs-\br} and the second law of
thermodynamics leads to a bound on the equation of state as eq. \cw.

Quantum cosmology is an elegant idea to probe the origin of our
universe, which says that our universe was born of a tunnelling from
nothing. Since the volume of the spatial flat or open universe is
infinite, unless the topology or the configuration of the universe
is nontrivial \fl, the tunneling probability from nothing to any one
of them is suppressed. Here we focus on the case with trivial
topology and only a closed universe can emerge. In quantum cosmology
scenario, the initial state corresponds to $H=\dot a/a=0$. Nothing
is interpreted as the initial state of our universe with zero
entropy \qh, since the entropy of a closed universe takes the form
as \entp.

The tunnelling probability corresponding to Hartle-Hawking wave
function of the universe is given by ${\cal P}_{HH} \simeq \exp
\left({3 \over 8 \Lambda} \right)$. A universe with cosmological
constant $\Lambda=0$ is favored. This universe is empty. It
contradicts the Hot big bang history of our universe. We expect that
the effects of quantum gravity can improve the wave function of the
universe to prefer a universe with matter within it when it was
born, for instance \refs{\fst-\hm, \qh}. Since the Hubble parameter
equals zero when the universe was created, the critical energy
density equals zero by using eq. \rc. Now $\Omega=-\Omega_k
\rightarrow +\infty$. Using eq. \cw, Holographic principle for a
closed universe requires the equation of state for the matter in the
initial state of our universe satisfy \eqn\wi{w_i\leq -1+{2 \over
n}. } Thus a universe filled with radiation or the dust-like matter
can not be created in the quantum cosmology scenario. If the
equation of state for the matter when the universe was born is
roughly a constant, the expansion of the universe must be
accelerating, or equivalently, an inflationary universe naturally
emerges. On the other hand, Wheeler-DeWitt (WdW) equation which
describes the evolution of the wave function of the universe $\Psi$
is \eqn\wdw{\left({\hat \Pi}_a^2+U(a)\right)\Psi=0, } where the
potential is $U(a)\sim a^{2n-4}-a^{2n-2}G\rho$, ${\hat \Pi}_a=-i\p
/\p a$ and the energy density for the matter satisfies $\rho \sim
a^{-n(w+1)}$. In order that there is a barrier in the potential
$U(a)$ and the universe can be created from nothing, the equation of
state for the matter should satisfy $w< -1+2/n$ which is consistent
with our above argument and prefers an inflationary universe. A
similar result in four dimensions has also been obtained in \ba.

In \br, Bak and Rey suggested to consider apparent horizon instead
of the particle horizon instead of the particle horizon in \fs\ and
they found a constraint on the equation of state as $w\leq -1+2/n$
in $(n+1)$ dimensions. This constraint means that the expansion of
the universe is always accelerating, which conflicts with the
history of our universe. But the bound suggested by us in eq. \cw\
is softer. When our universe was born, $\Omega=-\Omega_k\rightarrow
+\infty$ and $w \leq -1+2/n$. After a stage of inflation, the energy
density of the curvature was inflated away and $\Omega \rightarrow
1$. Now the constraint on the equation of state became $w \leq 1$,
which allows that the reheating occurred and the evolution of our
universe could be dominated by radiation or matter.

On the other hand, since the geometry of a universe can not be
changed classically, we can expect our universe is always closed in
the quantum cosmology scenario. If there are only radiation and
dust-like matter in such a closed universe, the bound on the
equation of state \cw\ will be violated sooner or later, unless
there is also dark energy with $w < -1+2/n$. For instance, we take a
closed universe dominated by dust-like matter into account. The
Friedman equation takes the form \eqn\dfe{H^2 = {16\pi G \over
n(n-1)} {\rho_0 \over a^n}-{1 \over a^2}, } with $a_0=1$. When the
scale factor goes to \eqn\amx{a=a_{max}=\left({16\pi G \rho_0 \over
n(n-1)} \right)^{1 \over n-2}, } Hubble parameter equals zero and
the universe begins to collapse. Now the bound \cw\ becomes $w\leq
-1+2/n$ and is violated. A more careful consideration tells us that
the holographic bound \cw\ has been violated before $a\rightarrow
a_{max}$. To avoid this problem, there should be a matter, named
dark energy, with equation of state $w<-1+2/n$. Otherwise the
universe will collapse and the holographic bound \cw\ is violated.
Therefore we says that the holographic principle predicts that there
must be dark energy today in our universe. In fact, the dark energy
has also been confirmed by the cosmological observation
\refs{\wmap,\ex} at a high level of statistical significance.

To summarize, we obtain a bound on the equation of state for the
matter in a flat or closed universe. Using this bound, we find
that an inflationary universe naturally emerges and today's dark
energy is also needed in the history of our universe in the
quantum cosmological scenario. The matching of the whole history
of our universe seems mysterious, but it can be taken as evidence
that the our proposal is on the right track. It is the first time
that the accelerated expansion is necessary from the fundamental
theory point of view. Since the initial entropy of our universe in
the quantum cosmology equals zero, we can also expect that the
entropy should increase with the classical evolution of our
universe. This naturally provides an arrow of time, i.e. along the
line with entropy increasing. Because Hubble parameter and the
entropy drops to zero when the closed universe begins to collapse,
our universe expands for ever; otherwise, the second law of the
thermodynamics will be violated.

Unfortunately, we do not understand the microscopic physics about
the deep correspondence between Friedman equation for a closed
universe and the formulation for the entropy of the CFT. The
physical meaning of the entropy formula \entp\ is also unknown. We
believe that this deep duality encodes some unknown, but important
insights on the holographic principle in a no boundary system, and
it may play a critical role on our understanding of the nature of
the spacetime. We also hope this work can open a window to
understand the history of our universe.

\bigskip

Acknowledgments

We would like to thank M. Li, J.X. Lu, J.H. She and P. Zhang for
useful discussions. We also thank E. Verlinde for encouragement.
This work was supported by a grant from NSFC, a grant from China
Postdoctoral Science Foundation and a grant from K. C. Wang
Postdoctoral Foundation.

\listrefs
\end